\documentclass[useAMS,usenatbib]{mn2e}
\usepackage{graphicx}

\title{A major outburst from the X-ray binary RX J0520.5-6932.}
\author[W. R. T. Edge et al.]
       {W. R. T. Edge, M. J. Coe, J. L. Galache, A. B. Hill \\
        School of Physics and Astronomy, Southampton University, SO17 1BJ}

\begin{document}

\date{Accepted 2003.
      Received 2003;
      in original form 2003}

\pagerange{\pageref{firstpage}--\pageref{lastpage}} \pubyear{2003}

\maketitle

\label{firstpage}

\begin{abstract}

We report on the analysis of 8 years of MAssive Compact Halo
Objects (MACHO) data for the source RX J0520.5-6932. A regular
period of 24.4 days has been confirmed, however this is manifest
almost entirely in the red part of the spectrum. A major outburst,
lasting approximately 200 days, was observed which increased the
apparent brightness of the object by approximately 0.15 magnitudes
without significantly altering its V-R colour index. This outburst
was also seen in X-ray data. The evidence from this analysis
points to the identification of this object as a Be/X-ray binary
with a periodically variable circumstellar disk and a very early
optical counterpart.

\end{abstract}


\section{INTRODUCTION}
\label{sec:introductionntificat}

\subsection{Be/X-ray Binaries}
\label{subsec:mylabel4}

Most High Mass X-ray Binaries (HMXBs) belong to the Be class, in
which a neutron star orbits an OB star surrounded by a
circumstellar disk of variable size and density (Negueruela and
Coe 2002).

The optical companion stars are early-type O-B class stars of
luminosity class III-V, typically of 10 to 20 solar masses that at
some time have shown emission in the Balmer series lines. The
systems as a whole exhibit significant excess flux at long (IR and
radio) wavelengths, referred to as the infrared excess. These
characteristic signatures as well as strong H$\alpha $ line
emission are attributed to the presence of circumstellar material
in a disk-like configuration (Coe 2000, Okazaki and Negueruela
2001).

The mechanisms which give rise to the disk are not well
understood, although fast rotation is likely to be an important
factor, and it is possible that non-radial pulsation and magnetic
loops may also play a part. The disk is thought to consist of
relatively cool material, which interacts periodically with a
compact object in an eccentric orbit, leading to regular X-ray
outbursts. It is also possible that the Be star undergoes a sudden
ejection of matter (Negueruela 1998).

Be/X-ray binaries can present differing states of X-ray activity
varying from persistent low or non-detectable luminosities to
short outbursts. Systems with wide orbits will tend to accrete
from less dense regions of the disk and hence show relatively
small outbursts. These are referred to as Type I and usually
coincide with the periastron of the neutron star. Systems with
smaller orbits are more likely to accrete from dense regions over
a range of orbital phases and give rise to very high luminosity
outbursts, although these may be modulated by the presence of a
density wave in the disk. Prolonged major outbursts, which do not
exhibit signs of orbital modulation, are normally called Type II
(Negueruela 1998).

\subsection{Previous observations of RX J0520.5-6932}

The X-ray source RX J0520.5-6932 was discovered by ROSAT on 11 Feb
1991 at a luminosity of $5\times10^{34}$ ergs/s and identified
with a $V\sim14$ magnitude star (Schmidtke et al, 1994). An
analysis of the Optical Gravitational Lensing Experiment (OGLE)
data by Coe et al. (2001) revealed the presence of a 24.45 day
periodic modulation of about 0.03 magnitudes. In the same paper a
spectral type of O9V was proposed for the optical counterpart.

\section{OPTICAL AND IR PHOTOMETRY}

In 1992 the  MAssive Compact Halo Objects project (MACHO) began a
survey of regular photometric measurements of several million
Magellanic Cloud and Galactic bulge stars (Alcock et al. 1993).

The MACHO data on RX J0520.5-6932 (Database identification number
78.6464.23) cover the period July 1992 to January 2000 and consist
of lightcurves in two colour bands described as \textit{blue} and
\textit{red}. \textit{Blue} is close to the standard $V$ passband
and \textit{Red }occupies a position in the spectrum about halfway
between $R$ and $I$ (Alcock et al. 1999). Transformation equations
are given in the same paper which enable these values to be
converted to $V$ and $R$ and, where appropriate, we have used the
latter, adopting the convention used by Schmidtke et al. (2003) to
refer to them as $V_{MAC}$ and $R_{MAC}$. However, because they
depend on a combination of both \textit{blue} and \textit{red
}values to obtain the conversions, and because, as we explain
later, we have found significant differences in the modulation of
the two lightcurves, we have used the original MACHO values for
the purposes of analysis. Out of a total of 1678 observations,
1546 were used, the remainder being rejected because of the value
of the errors (9.999 magnitudes).

\begin{figure}\begin{center}
\includegraphics[width=85mm]{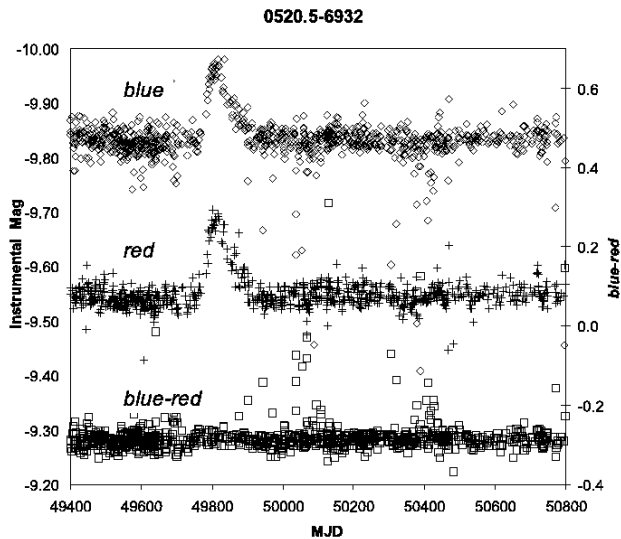}
\caption{RX J0520.5-6932 MACHO lightcurves. The whole dataset is
shown. The left vertical axis shows MACHO instrumental values. The
right one shows the \emph{Blue} minus \emph{Red} colour index.}
\label{fig:fig1}
\end{center}
\end{figure}

It was immediately evident that, in both colour bands, there was a
large isolated peak at about MJD 49810 (Apr 1995) extending over a
period of approximately 200 days (Figure~\ref {fig:fig1}). At its
maximum, this corresponded to an increase in magnitude of about
0.15 above the mean, or a 15{\%} change in flux. A \textit{blue}
minus \textit{red} colour index did not however reveal any
significant change in the relative brightness of the two colour
bands over the duration of the peak or throughout the rest of the
observations. No other comparable peak was found anywhere else in
this dataset or in the previously published OGLE data (Coe et al,
2001) which do not cover the date of this outburst.

\begin{figure}
\begin{center}
\includegraphics[width=85mm]{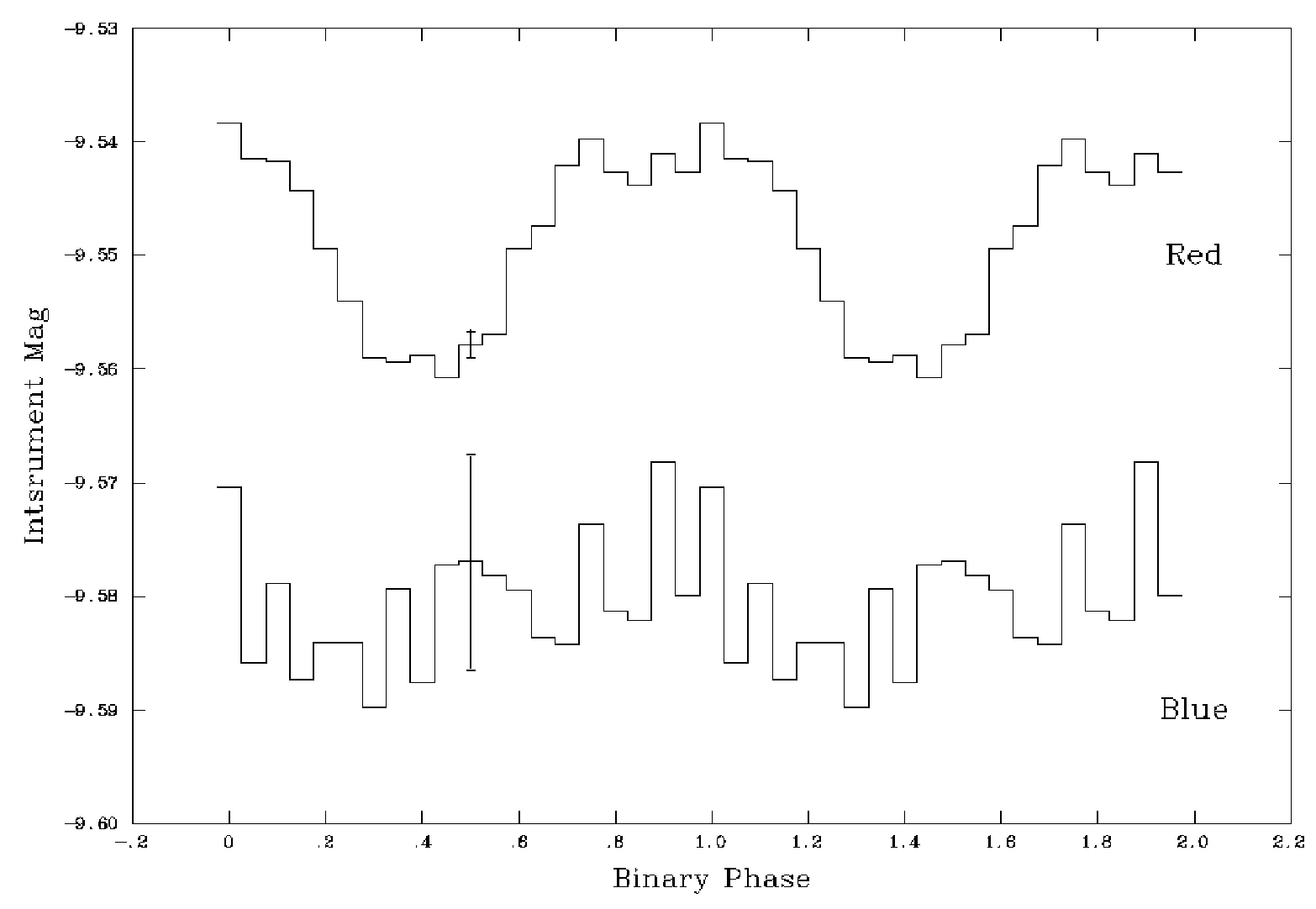}
\caption{The \emph{Red} and \emph{Blue} phase profiles folded at a
period of 24.4 days using the whole dataset. Typical error bars are shown.}
\label{fig:fig2}
\end{center}
\end{figure}

Both MACHO lightcurves were tested for periodicity using the
Starlink PERIOD Lomb-Scargle algorithm and a regular period of
24.44 $\pm$ 0.05 days was detected in the \textit{red} curve. A
consistent period could not, however, be detected when the
\textit{blue} curve was analysed using the whole dataset. A
comparison of the period profiles at 24.4 days is shown in
Figure~\ref {fig:fig2}. A weak period of 24.4 days was however
detected in the \textit{blue} curve during the 500 days
immediately preceding the outburst. The modulation of the red
curve amounted to a magnitude of 0.02 equating to a variation in
flux of 2{\%}. Coe et al. (2001) have detected a modulation in the
OGLE I band of 0.03 magnitudes or 3{\%} flux change.

The optical period was present during the outburst. The latter,
however, had a rather complex profile which could not reliably be
subtracted from the underlying continuum without distorting the
data, moreover it only covered about 7 cycles of the optical
period so that attempts to plot the profile of the 24.4 day period
did not yield meaningful results. Significant differences in the
characteristics of the modulation were not found when the data
were divided into three epochs, i.e: before the outburst (up to
MJD 49730), during the outburst (MJD 49730 - 49920) and after the
outburst (after MJD 49920).

\begin{figure}
\begin{center}
\includegraphics[width=85mm]{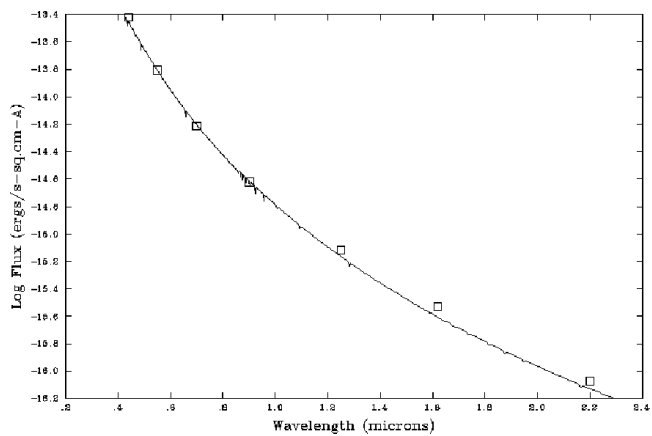}
\caption{Optical and IR magnitudes plotted against the model
atmosphere of an O9 star.} \label{fig:fig3}
\end{center}
\end{figure}

Using the STARLINK programme DIPSO, measured photometric
magnitudes were plotted against the spectrum of a model stellar
atmosphere (Kurucz, 1979). U, B, V and I were taken from average
OGLE values as published by Coe et al. (2001). The R value was
taken from direct measurements published in the same paper. IR
magnitudes were taken from the NASA/IPAC Infrared Science Archive
2MASS Second Incremental Release Point Source Catalog (PSC) (2000
March) (http://irsa.ipac.caltech.edu/) and have the following
values: J=14.4 $\pm$ 0.06, H=14.2 $\pm$ 0.07, K=14.3 $\pm$ 0.1.
These observations are dated 20 Dec 98 (MJD 51167). The combined
graph is shown in Figure~\ref {fig:fig3}. The model atmosphere
shown corresponds to a temperature of 31000 K (log g=4) and is
consistent with an O9 star. After normalising the model in the V
band and applying an E(B-V)=0.32 (Coe et al, 2000) there is a
small indication of an infra-red excess.

\section{OPTICAL SPECTROSCOPY}

\begin{figure}
\begin{center}
\includegraphics[width=85mm]{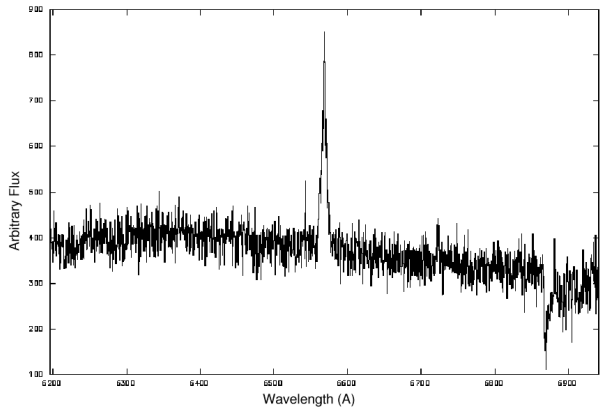}
\caption{$H\alpha$ spectrum of RX J00520.5-6932 taken at SAAO on
12 Nov 2001 (MJD 52225).} \label{fig:fig4}
\end{center}
\end{figure}

Spectroscopic observations in H$\alpha $ of the optical
counterpart to RX J0520.5-6932 were made with the SAAO 1.9m
telescope on 12 Nov 2001 (MJD 52225) and again on 14 Dec 2002 (MJD
52622). A 1200 lines mm$^{ - 1}$ reflection grating blazed at
6800{\AA} was used with the SITe CCD which is effectively 266 x
1798 pixels in size, creating a wavelength coverage of  6160{\AA}
to 6980{\AA}. The intrinsic resolution in this mode was
0.42{\AA}/pixel. The Nov 2001 spectrum is shown in Figure~\ref
{fig:fig4} and was measured to have an equivalent width of
$5.2\pm0.2${\AA}. By Dec 2002 the equivalent width had reduced to
$2.0\pm0.6${\AA}.

The Nov 2001 (MJD 52225) line is narrow and has a sharp single peak. The Dec
2002 (MJD 52622) line is less well defined and, because it is much weaker, a
clear shape cannot be discerned.

\section{X-RAY DATA}

In addition to the original X-ray detection in Feb 91 (MJD 48298), 2 further
ROSAT detections have been catalogued, on 05 Mar 92 (MJD 48686) and 24 Aug 93 
(MJD 49223)
(White et al. 2000). These showed a variation in flux of a factor
of 2 over the three measurements.

After the discovery of the optical peak in the MACHO data, an
analysis of the Burst And Transient Source Experiment data (BATSE)
was carried out over the same period. BATSE, on board the Compton
Gamma Ray Observatory, provided near continuous monitoring of the
whole sky during 1991-2000, in the hard X-ray band (Fishman et
al., 1989). The data types available from the 8 LADs (Large Area
Detectors) are the 16-channel continuous or "CONT" data, sampled
at 2.048 s intervals, and the 4-channel discriminator or "DISCLA"
data sampled every 1.024 s. The data presented in this paper make
use of the "CONT" data in the 20-70 keV range and further make use
of the Earth Occultation Technique. The EOT takes advantage of the
periodic eclipsing by the Earth of a source by measuring its total
flux twice in each ~93 minute orbit, once as it moves behind the
Earth's limb and then again when it reappears (Harmon et al.,
2002).

\begin{figure}
\begin{center}
\includegraphics[width=85mm]{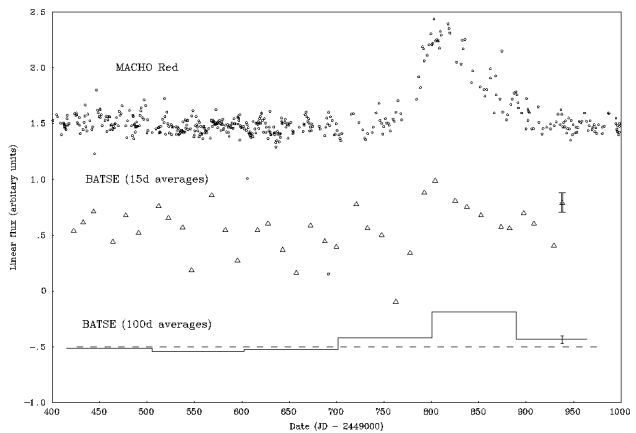}
\caption{The MACHO \emph{red} lightcurve plotted against BATSE
data on the same timescale. The middle curve is binned at 15 days
and the bottom curve at 100 days. Typical errors on the 15 and 100
day averages are indicated. The dotted line on the bottom curve shows the 
zero flux level. Both curves appear to confirm the
occurrence of an X-ray outburst at the same time as the optical
peak.} \label{fig:fig5}
\end{center}
\end{figure}

This analysis revealed a concurrent X-ray outburst with a similar
profile. This is shown, on the same timescale as the optical data,
in Figure~\ref {fig:fig5}. Signal to noise constraints determined
that the minimum binning duration had to be too long to permit
detection of the 24.4 day period in the BATSE data.

From the BATSE data we are able to determine the X-ray flux at the
outburst peak to be $\sim$20 mCrab. Assuming a distance of 50 kpc to
the LMC this implies an intrinsic X-ray luminosity of $\sim$8 x
$10^{38}$ erg/s. The significance of the outburst may be estimated
from the 100d BATSE samples shown in Figure~\ref {fig:fig5} and is
found to be approximately 13 sigma. The flux was not detectable
outside the outburst.

\section{DISCUSSION}

The observational evidence summarised above supports the
identification of this variable X-ray source with a Be/X-ray
transient. The detection of a prominent H$\alpha $ line in the
spectrum points strongly to the presence of a circumstellar disk,
indicating a Be X-ray binary, although the line shows no
indication of a broadened double peak which might indicate a
rapidly rotating disk viewed from an angle close to the plane of
rotation. It can be concluded from this, and from the relatively
small value of the H$\alpha $ equivalent width, that the system is
being observed from an angle close to the axis of rotation and
that the circumstellar disk is small. This is supported by the relatively small
value of the observed infra-red excess in the Dec 98 (MJD 51167)
2MASS data. The greatly diminished H$\alpha $ equivalent width in
Dec 2000, as compared with the previous year, further suggests a
highly variable disk regime. The well-established relationship
between the $H\alpha$ EW and intrinsic (J-K) can be used to ensure
we have agreement on the small size of the circumstellar disk (Coe
et al, 1993). Using E(B-V)=0.32 we determine $(J-K)_{o} = -0.1$
and we have measured the $H\alpha$ EW to be in the range 2-5\AA.
Comparing these values with the correlation diagram shown in
Figure 4 of Coe et al. (1993) clearly indicates that we are dealing
with a Be system with an unusually small disk and our values agree
with the general trend of the data in this figure.

The isolated optical flare seen in the MACHO data appears to be
similar in size and duration to those identified as type-1 Be star
candidates in the Small Magellanic Cloud by Mennickent et al.
(2002). As such the rise in luminosity is evidence for an outburst
although it is somewhat surprising in that it is not accompanied
by any significant change in the \textit{blue} minus \textit{red}
colour index (unlike the stars in Mennickent et al. which tend to
be redder when brighter). The discovery of a concurrent X-ray peak
in the BATSE data lends support to the conjecture that this event
signals a major convulsion in the optical star, or its
circumstellar disk, leading to high mass accretion and X-ray
emission by its neutron star companion. It should be mentioned in
passing that an isolated giant X-ray outburst would be described
as a Type-II outburst by contrast with the optical classification
used earlier.

The 24.4 day optical modulation in the \textit{red} MACHO data,
which was also present in the OGLE I data, may represent the
binary period of the system. The almost complete absence of
modulation in the blue curve suggests that this effect is
occurring in the circumstellar disk and could be explained if a
periodic distortion were being occasioned by the eccentricity of
the neutron star orbit. It is possible that the latter is, in
fact, causing the disk to alternate between resonant states
(Okazaki et al., 2002).

Two other Magellanic Cloud systems with modulated optical
lightcurves have already been reported by Charles et al. (1983)
for A0538-66 and Schmidtke et al. (2003) for RX J0058.2-7231. It
would therefore seem that this is not an isolated phenomenon but
may be representative of an optically variability typical of some
Be/X-ray binaries.

It might be anticipated that, if the modulation in the \emph{red}
curve were solely attributable to the circumstellar disk, it would
not be present around the time of 2MASS observations (20 Dec 98 -
MJD 51167) when any evidence for infra-red excess was not
observed. Unfortunately this is difficult to test because there is
a gap in the MACHO data immediately prior to that date and also
the timespan required to properly observe a period (about 5
cycles) would be long enough for the disk to have reappeared. We
note also that the depth of modulation (2-3{\%}) is well within
the errors of the IR magnitudes. The relatively low infra-red
excess observed in Figure~\ref {fig:fig3} may simply indicate that
the disk is very small and the variation in the $H\alpha$
equivalent width is probably due to its changing size.

Using theoretical and empirical evidence we should be able to
predict the neutron star spin frequency from its orbital period
(Corbet et al., 1999). A pulsar in an orbit of 24.4 days would
thus be expected to have a pulse period of about 3-4 seconds.
However, so far, no convincing detection at this frequency has
been associated with this object.

From the X-ray and optical data presented here it is possible to
determine values for the Lx/Lopt ratio both during, and outside
the outburst. The optical luminosity is integrated over the range
400 -700 nm and is calculated assuming that the R band flux quoted
in Coe et al (2001) is valid during quiescence and is
representative of the whole optical range. For the outburst it
assumed that the optical brightness increased by the 0.15
magnitudes reported here. For the X-ray, the BATSE luminosity
during the outburst of $\sim$20 mCrab was used in conjunction with
an assumed Crab-like spectrum integrated over the range 2 - 100
keV. For the quiescent state a 3$\sigma$ upper limit of
$\le$10mCrab was determined. Using these values the quiescent
Lx/Lopt is found to be $\le$100, and the outburst Lx/Lopt to be
$\sim$230. Though these values are at towards the top end for a
HMXB system (Bradt \& McClintock, 1983) the event reported here is
undoubtably a substantial Type II outburst.

The identification of this system with an O9 star, if it is
correct, represents the discovery of a Be X-ray binary with a
counterpart of an unusually, although not uniquely,  early
spectral type. Further study may determine whether this has any
bearing on the isolated nature of the optical/X-ray outburst and
the evident variability of the circumstellar disk.

\section*{Acknowledgments}

This paper utilizes public domain data obtained by the MACHO
Project, jointly funded by the US Department of Energy through the
University of California, Lawrence Livermore National Laboratory
under contract No. W-7405-Eng-48, by the National Science
Foundation through the Center for Particle Astrophysics of the
University of California under cooperative agreement AST-8809616,
and by the Mount Stromlo and Siding Spring Observatory, part of
the Australian National University. This research has also made
use of the NASA/ IPAC Infrared Science Archive, which is operated
by the Jet Propulsion Laboratory, California Institute of
Technology, under contract with the National Aeronautics and Space
Administration, and the Optical Gravitational Lensing Experiment
(OGLE) Stellar Photometric Maps of the Small Magellanic Cloud,
through the Princeton University site. We are grateful to the
staff of SAAO for their support during our observations.


\newpage

\begin{thebibliography}{99}

\newcommand{\mnras}{MNRAS}
\newcommand{\aap}{A\&A}
\newcommand{\aaps}{A\&AS}
\newcommand{\apss}{Ap\&SS}
\newcommand{\aj}{AJ}
\newcommand{\apjs}{ApJS}
\newcommand{\iaucirc}{IAUC}
\newcommand{\pasj}{PASJ}
\newcommand{\pasp}{PASP}
\newcommand{\apj}{APJ}
\newcommand{\nyp}{Not yet published}


\bibitem[Alcock et al.(1999)]{1999PASP..111.1539A} Alcock, C.~et al.\ 1999, \pasp, 111, 1539
\bibitem[Alcock et al.(1993)]{1993sspp.conf..291A} Alcock, C.~et al.\ 1993, ASP Conf.~Ser.~ 43: Sky Surveys.~Protostars to Protogalaxies, 291
\bibitem[Charles et al.(1983)]{1983MNRAS.202..657C} Charles, P.~A.~et al.\ 1983, 
\bibitem[Coe et al. (1993)]{}Coe, M.~J., Everall, C., Norton, A.~J, Roche, P., Unger, S.~J, Fabregat, J., Reglero, V., \& Grunsfeld, J.~M.,\ 1993, \mnras, 261, 599.
\bibitem[Coe(2000)]{2000bpet.conf..656C} Coe, M.~J.\ 2000, ASP Conf.~Ser.~214: The Be Phenomenon in Early-Type Stars, 656
\bibitem[Coe et al.(2001)]{2001MNRAS.324..623C} Coe, M.~J., Negueruela, I., Buckley, D.~A.~H., Haigh, N.~J., \& Laycock, S.~G.~T.\ 2001, \mnras, 324, 623
\bibitem[Corbet, Marshall, Peele, \& Takeshima(1999)]{1999ApJ...517..956C} Corbet, R.~H.~D., Marshall, F.~E., Peele, A.~G., \& Takeshima, T.\ 1999, \apj, 517, 956
\bibitem[]{} Corbet, R.H.D., Marshall, F.E., Coe, M.J., Laycock, S., {\&} Handler, G. 2001, AJ, 548, L41
\bibitem[]{}Fishman et al. 1989, Proc. GRO Science Workshop, ed W.N. Johnson, NASA pub., 2-39
\bibitem[]{}Harmon, B. A., Fishman, G. J., Wilson, C. A., Paciesas, W. S., Zhang, S. N., Finger, M. H., Koshut, T. M., McCollough, M. L., Robinson, C. R., Rubin, B. C. 2002, ApJS, 138, 149
\bibitem[Kurucz(1979)]{1979ApJS...40....1K} Kurucz, R.~L.\ 1979, \apjs, 40,1
\bibitem[]{}Mennickent, R. E., Pietrzy{\'n}ski, G., Gieren, W., \& Szewczyk, O.\ 2002, \aap, 393, 887
\bibitem[]{} Negueruela, I.~\& Coe, M.~J.\ 2002, \aap, 385, 517
\bibitem[]{} Negueruela, I.\ 1998, \aap, 338, 505-510
\bibitem[]{}Okazaki, A.~T.\ 1993, \apss, 210, 369
\bibitem[]{} Okazaki, A.T. {\&} Negueruela, I. 2001, A{\&}A, 377, 161
\bibitem[Okazaki, Bate, Ogilvie, \& Pringle(2002)]{2002MNRAS.337..967O} Okazaki, A.~T., Bate, M.~R., Ogilvie, G.~I., \& Pringle, J.~E.\ 2002, \mnras, 337, 967
\bibitem[Schmidtke et al.(1994)]{1994PASP..106..843S} Schmidtke, P.~C., Cowley, A.~P., Frattare, L.~M., McGrath, T.~K., Hutchings, J.~B., \& Crampton, D.\ 1994, \pasp, 106, 843
\bibitem[Schmidtke et al.(2003)]{Not yet published} Schmidtke, P.~C., Cowley, A.~P., Levenson, L\ 2003, Not yet published
\bibitem[Schmidtke et al.(1999)]{1999AJ....117..927S} Schmidtke, P.~C., Cowley, A.~P., Crane, J.~D., Taylor, V.~A., McGrath, T.~K., Hutchings, J.~B., \& Crampton, D.\ 1999, \aj, 117, 927
\bibitem[White, Giommi, \& Angelini(2000)]{2000yCat.9031....0W} White, N.~E., Giommi, P., \& Angelini, L.\ 2000, VizieR Online Data Catalog, 9031,


\end{thebibliography}
\end{document}